\documentclass[twocolumn,showpacs,amsmath,amssymb, nobibnotes, aps, prl,showkeys]{revtex4}
\usepackage{graphicx}% Include figure files
\usepackage{dcolumn}% Align table columns on decimal point
\usepackage{bm}% bold math
\usepackage{docs}

\expandafter\ifx\csname package@font\endcsname\relax\else
 \expandafter\expandafter
 \expandafter\usepackage
 \expandafter\expandafter
 \expandafter{\csname package@font\endcsname}%
\fi

\begin{document}

\title{Describing correlated observations of neutrino and gamma ray flares from the blazar TXS 0506+056 with proton blazar model}

%\title{Interpretation of the coincident observation of high energy gamma rays and a neutrino event associated with the flaring blazar TXS 0506+056 with hadronic interaction model}

\author{Prabir Banik$^{*}$\thanks{Email address: pbanik74@yahoo.com} and Arunava Bhadra$^{\dagger}$\thanks{Email address: aru\_bhadra@yahoo.com}}
\affiliation{ $^*$Surendra Institute of Engineering $\&$ Management, Dhukuria, Siliguri, West Bengal, India 734009\\}

\affiliation{ $^{\dagger}$High Energy $\&$ Cosmic Ray Research Centre, University of North Bengal, Siliguri, West Bengal, India 734013\\}
% and Debasish Majumdar$^3$\thanks{Email address: debasish.majumdar@saha.ac.in}
%\affiliation{ $3$ Astroparticle Physics and Cosmology Division, Saha Institute of Nuclear Physics, 1/AF Bidhannagar, Kolkata, West Bengal, India 700064\\}
\begin{abstract}

Recent detection of the neutrino event, IceCube-170922A by IceCube observatory from the Blazar TXS 0506+056 in the state of enhanced gamma ray emission indicates for acceleration of cosmic rays in the blazar jet. The non-detection of the broadline emission in the optical spectrum of TXS 0506+056 and other BL Lac objects suggests that external photons emissions are weak and hence photo-meson ($p\gamma$) interaction may not be a favored mechanism for high energy neutrino production. The lack of broadline signatures also creates doubt about the presence of a high density cloud in the vicinity of the super-massive black hole (SMBH) of TXS 0506+056 and consequently raised question on hadronuclear ($pp$) interaction interpretation like relativistic jet meets with high density cloud. Here we demonstrate that non-relativistic protons in the proton blazar model, those come into existence under charge neutrality condition of the blazar jet, offer sufficient target matter for $pp-$interaction with shock accelerated protons and consequently the model can describe consistently the observed high energy gamma rays and neutrino signal from the blazar TXS 0506+056. 

\end{abstract}

\maketitle

\section{Introduction}
Very recently IceCube Neutrino Observatory reported the detection of a high-energy muon neutrino event, IceCube-170922A of energy $\sim 290$ TeV with a 56.5\% probability of being a truly astrophysical neutrino \cite{IceCube18a, IceCube18b}. The best-fit reconstructed arrival direction of the neutrino was consistent with the $0.1^0$ from the sky location of a flaring gamma ray blazar TXS 0506+056 \cite{IceCube18a, Ansoldi18}. As a follow-up observation Fermi Large Area Telescope (LAT) Collaboration \cite{Tanaka17} reported that the direction of origin of IceCube-170922A was consistent with the known gamma ray source TXS 0506+056 blazar which was in a state of enhanced emission with day scale variability \cite{Keivani18} on 28 September 2017. The observed association of a high-energy neutrino with a blazar during a period of enhanced gamma-ray emission suggests that blazars may indeed be one of the long-sought sources of very-high-energy cosmic rays and hence these observations offers a unique possibility to explore the interrelate between energetic gamma rays, neutrinos, and cosmic rays.

The electromagnetic spectral energy distribution (SED) of the blazar TXS 0506+056 exhibits a double-hump structure which is a common feature of the non-thermal emission from blazars. The first hump, which peaking in the optical-ultraviolet range, is usually attributed to synchrotron radiation and the higher energy hump with peak energy in the GeV range, is often interpreted due to inverse Compton (IC) emission. An archival study of the time-dependent $\gamma$-ray data over the last ten years or so reveal that the source was in quiescent stage most of the time, the flaring was noticed during the period July 2017 to September 2017. The averaged integrated flux above 0.1 GeV from TXS 0506+056 was found $(7.6 \pm 0.2) \times 10^{-8}$ $cm^{-2}s^{-1}$ during 2008 to 2017 from Fermi-LAT observations which in the week 4 to 11 July 2017 elevates to the level $(5.3 \pm 0.6) \times 10^{-7}$ $cm^{-2}s^{-1}$. The Astro-Rivelatore Gamma a Immagini Leggero (AGILE) gamma-ray telescope obtained flux of $(5.3 \pm 2.1)\times 10^{-7}$ $cm^{-2}s^{-1}$ during 10 to 23 September 2017. The Major Atmospheric Gamma Imaging Cherenkov (MAGIC) Telescopes detected a significant very-high-energy $\gamma$-ray signal with observed energies up to about 400 GeV on 28 September 2017. Note that the Icecube observatory detected the neutrino event on 22 September 2017. It was found from optical to x-ray observations that the lower energy hump of the SED of the source does not show any noticeable time variation over the stated period of study.    

Several efforts have been made so far to model the production of the detected neutrino event together with the electromagnetic (EM) observations from TXS 0506+056. Mainly two different production scenario, namely lepto-hadronic ($p\gamma$) \cite{Ansoldi18,Keivani18,Gao18,Cerruti19} and hadronic (pp) \cite{Liu18,Sahakyan18} have been proposed in the literature to interpret the observations. A common feature of all the proposed models is that protons, like electrons, are also assumed to be accelerated to relativistic energies in the acceleration sites. Subsequently the accelerated protons interacting with low energy photons of blazar environment (lepto-hadronic interaction) and/or with ambient matter produce high energy gamma rays and neutrinos.  

Ansoldi et al. (2018) \cite{Ansoldi18} showed that the measured neutrino event from the said blazar can be interpreted consistently with the EM observations by assuming a dense field of external low-energy photons originating outside of the jet as targets for photohadronic interactions. The lack of broadline signatures in the optical spectrum of TXS 0506+056 and other BL Lac objects suggest that such external photons emissions may be weak \cite{Keivani18}. The model discussed in Ansoldi et al. (2018) \cite{Ansoldi18}, however, does not invoke radiation from broadlines but, instead, assume the existence of soft radiation produced in a possible layer surrounding the jet. Therefore the lack of broadlines does not impact this specific scenario. In this context it is also to be noted that the BL Lac nature of TXS 0506+056 has been recently questioned by Padovani et al. 2019 \cite{Padovani19}. Keivani et al. (2018) \cite{Keivani18} considered a hybrid leptonic scenario of TXS 0506+056 where the production of high energy gamma rays was interpreted by external inverse-Compton processes and high-energy neutrinos via a radiatively sub-dominant hadronic component. 

For efficient high energy $\gamma-$ray production in AGN jet via $pp$ interaction demands high thermal plasma density; the thermal plasma in the jet should exceed $10^6$ cm$^{−3}$ in order to interpret the reported TeV flares of Markarian 501 by $pp$ interactions for any reasonable acceleration power of protons $L_p \le 10^{45}$ erg/s \cite{Aharonian00}. The stated pure hadronic mechanism thus can be effectively realized in a scenario like ``relativistic jet meets target" \cite{Morrison84}, i.e. considering that $\gamma-$radiation is produced in dense gas clouds that move across the jet \cite{Dar97}. Recently, Liu et al. (2018) \cite{Liu18} described the observed gamma ray \& neutrino flux from the blazar TXS 0506+056 by assuming the presence of clouds in the vicinity of the super-massive black hole (SMBH) that provides targets for inelastic $pp$ collisions once they enter the jet. Liu et al. (2018) considered the synchrotron emission and inverse Compton emission of secondary electrons produced in cascade when high energy $\gamma-$rays absorbed in $\gamma\gamma$ pair production with the emission region of the jet. However, the presence of broadline region (BLR) clouds in the vicinity of the SMBH for TXS 0506+056 is questionable due to the non-detection of the BLR emission from TXS 0506+056 and other BL Lac objects \cite{Keivani18}. 

The composition of bulk of the jet medium is not clearly known which makes difficulties to understand the interaction mechanism for gamma ray and neutrino production. But on average, jet plasma must be neutral to remain collimated \cite{Hirotani05}. Therefore, the two main scenarios for their matter composition are suggested: a `pair plasma' consisting of only of relativistic electrons and positrons \cite{Kino04} and a `normal plasma' consisting of (relativistic or non-relativistic) protons and relativistic electrons \cite{Celotti93}. A useful quantity that can furnish some constraints on jet composition is the kinetic power of an AGN jet. By comparing the bulk kinetic energy of the parsec scale jet with the kinetic luminosities on extended scales \cite{Rawlings91}, Celotti \& Fabian (1993) \cite{Celotti93} argued in favor of an electron-proton fluid. For high luminous blazars, to maintain the radiated power which would not exceed that carried by the jet, the proton component of plasma is necessary (see Ghisellini, 2010 \cite{Ghisellini10}, and references therein). 

Under the context, in the present work we exploit the main essences of proton blazar model \cite{Protheroe01, Mucke01} to explain the observed higher-energy bump of the EM SED along with the neutrino from the blazar TXS 0506+056 at flaring stage. The detected lower-energy bump of EM SED from the blazar can be well interpreted with the synchrotron radiation of relativistic electrons present in jet plasma whereas the cold (non-relativistic) protons density that arose from charge neutrality condition can provide sufficient target matter (proton) for production of high energy gamma rays and neutrinos via the $pp$ interaction. For TXS 0506+056 such a scenario is more realistic then the scenario like the cloud-in-jet model \cite{Aharonian17} as we argue later. We would also like to examine the maximum energy that a cosmic ray particle can attain in the blazar jet; the detected $\sim 290$ TeV energy neutrino alone suggest that acceleration of protons in the jet of this object to energies of at least several times $10^{15}$ eV. 

The organization of this paper is as follows: In the next section, we shall describe the methodology for evaluating the gamma-ray and neutrino  fluxes generated in interaction of cosmic rays with the ambient matter in the AGN jet under the framework of proton blazar model. The numerical results of the hadronically produced gamma-rays and neutrino fluxes from the AGN jet over the GeV to TeV energy range are shown in section III. The findings are compared with the observed gamma rays spectra and the neutrino event from the blazar and the results are discussed in the same section. Finally we conclude in Sec. IV.

\section{Methodology}

The overall jet composition of AGN is not properly known. In the adopted proton blazar inspired model it is assumed that the relativistic jet material is composed of relativistic protons (p) and electrons (e$^−$). Some cold protons also exist, allowing charge neutrality to be fulfilled. The ratio of number of relativistic protons to electrons, the maximum energies attained by protons/electrons in acceleration process and slope of their energy spectrum, luminosities of electrons and protons are adjustable parameters of the model. In this model flaring is produced due to high magnetic activities in the source (similar to the origin of flaring activities in the Sun). 

We consider a spherical blob of size $R_b'$ (primed variables for jet frame) in the AGN jet which is the region responsible for the blazar emission. The blob is moving with a Doppler factor $\delta = \Gamma_j^{-1}(1-\beta_j\cos\theta)^{-1}$ where $\theta$ is the angle between the line of sight and the jet axis and $\Gamma_j = 1/\sqrt{1-\beta_j^2}$ is the bulk Lorentz factor \cite{Petropoulou15}, and it contains a tangled magnetic field of strength $B'$.  

In the proton blazar framework the low-energy bump of the SED is explained by synchrotron radiation of accelerated relativistic electron in blazar jet having broken power law energy distribution as \cite{Katarzynski01}

\begin{eqnarray}
N_e'(\gamma_e') = K_e \gamma_e'^{-\alpha_1} \hspace{1.5cm} \mbox{if}\hspace{0.6cm} \gamma_{e,min}' \le \gamma_e' \le \gamma_b' \nonumber \\
         = K_e \gamma_b'^{\alpha_2-\alpha_1} \gamma_e'^{-\alpha_2} \hspace{0.35cm} \mbox{if}\hspace{0.45cm} \gamma_b' <\gamma_e' \le \gamma_{e,max}'\;
\end{eqnarray}
where $\gamma_e' = E_e'/m_e c^2$ is  the  Lorentz  factor of electrons of energy $E_e'$, $\alpha_1$ and $\alpha_2$ are the spectral indices before and after the spectral break Lorentz factor $\gamma_b'$ respectively. The normalization constant $k_e$ can be found from \cite{Bottcher13}
\begin{equation}
L_e' = \pi R_b'^2 \beta_j c \int_{\gamma_{e,min}'}^{\gamma'_{e,max}}m_e c^2\gamma_e' N_e(\gamma_e') d\gamma_e'
\end{equation}
where $L_e'$ is the kinetic power in relativistic electrons in the blazar jet frame. The number density of highly relativistic (`hot') electrons is $n_{e,h}' = \int N_e'(\gamma_e') d\gamma_e'$ and the corresponding energy density is $u_e' = 3p_e' = \int m_e c^2\gamma_e' N_e'(\gamma_e') d\gamma_e'$ where $p_e'$ is the radiation pressure due to relativistic electrons. Due to strong synchrotron and Inverse Compton cooling at relativistic energies, the acceleration efficiency of electrons in AGN jet is quite low and it can be assumed to be $\chi_e \approx 10^{-3}$ \cite{Bykov96,Eichler05,Vazza15}. Hence total number can be determined as $n_{e}' = n_{e,h}'/\chi_e$. Thus the number density of non-relativistic (`cold') electrons is given by $n_{e,c}' = n_{e}' - n_{e,h}'$.

The emissivity of photons of energy $E_{s}'$ ($= m_e c^2 \epsilon_{s}'$) due to the synchrotron emission of electrons which describe low energy component of EM SED of the blazar, can be written as \cite{Bottcher13}
\begin{eqnarray}
Q'_{s}(\epsilon_{s}') = A_0 \epsilon_{s}'^{-3/2} \int_1^\infty d\gamma_e' N_e'(\gamma_e')\gamma_e'^{-2/3}e^{-\epsilon_{s}'/(b\gamma_e'^2)}
\end{eqnarray}
with the normalization constant
\begin{eqnarray}
A_0 = \frac{c \sigma_T B'^2}{6\pi m_e c^2 \Gamma(4/3) b^{4/3}}, \nonumber
\end{eqnarray}
where $\sigma_T$ is the Thomson cross-section, $b = B'/B_{crit}$ and $B_{crit} = 4.4\times 10^{13}$ G. The magnetic field energy density is $u_B' = B'^2/8\pi = 3 p_B'$ where $p_B'$ is the corresponding pressure.

The emissivity of photons of energy $E_{c}'$ ($= m_e c^2 \epsilon_{c}'$) due to the inverse compton scattering of primary accelerated electrons
with the seed photons co-moving with the AGN jet, which can describe lower part of high energy component of EM SED of the blazar, can be written as\cite{Blumenthal70,Inoue96}
%The emissivity of photons of energy $E_{c}'$ ($= m_e c^2 \epsilon_{c}'$) due to the self-Compton radiation (SSC) of electrons \cite{Blumenthal70,Inoue96} which can describe lower part of high energy component of EM SED of the blazar, can be written as
%\begin{eqnarray}
%Q_{c}(\epsilon_{c}') = \int_0^\infty d\epsilon_{s}' n_s'(\epsilon_{s}') \int_{\gamma_{e,0}'}^{\gamma_{e,max}'} d\gamma' N_e'(\gamma_e') C(\epsilon_{c}',\gamma_e',\epsilon_{s}'),
%\end{eqnarray}

\begin{eqnarray}
Q_{c}(\epsilon_{c}') = \int_0^\infty d\epsilon_{j}' n_j'(\epsilon_{j}') \int_{\gamma_{e,0}'}^{\gamma_{e,max}'} d\gamma_e' N_e'(\gamma_e') C(\epsilon_{c}',\gamma_e',\epsilon_{j}'),
\end{eqnarray}
where $\gamma_{e,0}' = \frac{1}{2}\epsilon_{c}'\left(1+\sqrt{1+\frac{1}{\epsilon_{c}'\epsilon_{j}'}}\right)$ and the compton kernel $C(\epsilon_{c}',\gamma_e',\epsilon_{j}')$ is given by Jones (1968) \cite{Jones68} as
\begin{eqnarray}
C(\epsilon_{c}',\gamma_e',\epsilon_{j}') = \frac{2\pi r_e^2 c}{\gamma_e'^2\epsilon_{j}'} \Big[2k \ln(k) + (1+2k)(1-k)   \big.  \nonumber   \\
\big. + \frac{(4\epsilon_{j}'\gamma_e' k)^2}{2(1+4\epsilon_{j}'\gamma_e'k)}(1-k)\Big],
\end{eqnarray}
with $k = \frac{\epsilon_{c}'}{4\epsilon_{j}'\gamma_e'(\gamma_e' - \epsilon_{c}')}$ and $r_e$ is the classical electron radius. Here, $n_{j}'(\epsilon_{j}')$ is the average number density of the seed photons of energy $\epsilon_{j}'$( in $m_ec^2$) in the blob of AGN jet which can be directly related to observed photon flux $f_{\epsilon_j}$ (in erg cm$^{-2}$ s$^{-1}$) from the blazar through \cite{Dermer02}
\begin{eqnarray}
\epsilon_{j}'n_{j}'(\epsilon_{j}') = \frac{2 d_L^2}{c R_b'^2\delta^2\Gamma_j^2} \frac{f_{\epsilon_j}}{m_e c^2\epsilon_{j}' }
\end{eqnarray}
where $\epsilon_{j} = \delta \epsilon_{j}'/(1+z) $ \cite{Atoyan03} relates photon energies in the observer and co-moving jet frame of red shift parameter $z$ respectively,  and $d_L$ is the luminosity distance of the AGN from the Earth. 

In the proton blazar model the cosmic ray protons are also supposed to accelerate to very high energies $E_p' = m_p c^2 \gamma_p'$ in the same region of blazar jet and the production  spectrum shall follow a power law \cite{Malkov01,Cerruti15}:  
\begin{equation}
 N_p'(\gamma'_p) =  K_p {\gamma'_p}^{-\alpha_p} .
\end{equation}
where $\alpha_p$ is the spectral index, $\gamma_p'$ is the Lorentz factor of accelerated protons, $K_p$ denotes the proportionality constant which can be found from the same expression as eq (2) but for protons and $L_p'$ is the corresponding jet power in relativistic protons. The number density of relativistic protons is $n_p' = \int N_p'(\gamma_p') d\gamma_p' $ and the corresponding energy density is $u_p' = 3p_p' = \int m_p c^2\gamma_p' N_p'(\gamma_p') d\gamma_p'$, where $p_p'$ is the radiation pressure due to relativistic protons. 

We estimate the mechanical luminosity or total kinematic jet power of an AGN jet containing jet frame energy density $u'$ (sum of $u_e'$, $u_p'$ and $u_B'$), pressure $p'$ (sum of $p_e'$, $p_p'$ and $p_B'$) and matter density $\rho'$ (including cold protons and electrons) from the following relation \cite{Protheroe01}
\begin{eqnarray}
L_{jet} = \Gamma_j^2 \beta_j c \pi R_b'^2\left[\rho'c^2(\Gamma_j-1)/\Gamma_j + u' + p' \right].
\end{eqnarray}
where we assume the Lorentz factor to be $\Gamma \approx \delta/2$ which is quite reasonable particularly for jets closely aligned to the line of sight of the observer. Applying charge conservation and considering that the number of relativistic electrons will be greater then the number of relativistic protons, the number of `cold' (non-relativistic) protons will be equal to the total number of electrons ($n_{e}'$) minus the number of hot protons ($n_p'$). Thus the cold matter density in protons and in electrons in the blob will be $\rho_p' = (n_{e}'-n_p')m_p$ where $m_p$ is the rest mass of a proton and $\rho_e' = n_{e,c}' m_e$ respectively. 

When the shock accelerated cosmic rays interact with the cold matter (protons) of density $n_{H} = \rho_p'/ m_p$ in the blob of AGN jet, the emissivity of produced secondary particles of energy $E_i' = m_e c^2 \epsilon_i'$ in co-moving AGN jet frame is given by \cite{Liu18,Anchordoqui07,Banik17a,Kelner06}
\begin{eqnarray}
Q_{i,pp}'(\epsilon'_{i}) = \frac{c n_{H}m_e}{m_p} \int_{\frac{m_e\epsilon'_{i}}{m_p}} \sigma_{pp}(E'_{p}) N'_p(\gamma'_{p})F_i\Big(\frac{E'_{i}}{E'_{p}},E'_{p}\Big)\frac{d\gamma'_{p}}{\gamma'_{p}}
\end{eqnarray}
where $i$ could be $\pi^{0}$ mesons, electrons (positrons) $e^{\pm}$ or neutrinos $\nu$ and $F_i$ is the spectrum of the corresponding secondary particles in a single $pp$ collision as given in Kelner et al. (2006) \cite{Kelner06}.

Due to decay of $\pi^0$ mesons, the resulting gamma ray emissivity as a function of gamma ray energy $E_{\gamma}'( = m_e c^2 \epsilon_{\gamma}')$ is given by \cite{Banik17b} 

\begin{eqnarray}
%\small
Q_{\gamma,pp}'(\epsilon'_{\gamma}) = 2\int_{\epsilon_{\pi,min}'(\epsilon'_{\gamma})}^{\epsilon_{\pi,max}'}\frac{Q_{\pi,pp}'(\epsilon'_{\pi})}{\left[{\epsilon'_{\pi}}^2-(\frac{m_{\pi}}{m_e})^2\right]^{1/2}}d\epsilon'_{\pi}
\end{eqnarray}
where ${\epsilon'_{\pi,min}}(\epsilon'_{\gamma}) = \epsilon'_{\gamma} + (\frac{m_{\pi}}{m_e})^2/(4\epsilon'_{\gamma})$ is the minimum energy of a pion required to produce a gamma ray photon of energy $\epsilon'_{\gamma}$ (in $m_e c^2$).

When propagating through an isotropic source of low-frequency radiation, the TeV$-$PeV gamma-rays can be absorbed at photon-photon ($\gamma\gamma$) interactions \cite{Aharonian08}. Thus, the emissivity of escaped gamma rays after $\gamma\gamma$-interaction can be written as \cite{Bottcher13} 
\begin{eqnarray}
Q_{\gamma,esc}'(\epsilon_{\gamma}') = Q_{\gamma}'(\epsilon_{\gamma}') .\left( \frac{1-e^{-\tau_{\gamma \gamma}}}{\tau_{\gamma \gamma}} \right).
\end{eqnarray}
Here $\tau_{\gamma \gamma}(\epsilon_{\gamma}')$ is the optical depth for the interaction and is given by \cite{Aharonian08}
\begin{eqnarray}
%\small
\tau_{\gamma \gamma}(\epsilon_{\gamma}') = R_b' \int \sigma_{\gamma\gamma}(\epsilon_{\gamma}',\epsilon_{j}') n_{j}'(\epsilon_{j}')d\epsilon_{j}'
\end{eqnarray}
where $\sigma_{\gamma\gamma}$ is the the total cross-section as given in Aharonian et al., 2008 \cite{Aharonian08} and $n_{j}'(\epsilon_{j}')$ describes the spectral distributions of target photons.  $n_{j}'(\epsilon_{j}')$ is generally assumed to be the observed synchrotron radiation photons produced by the relativistic electron population in co-moving jet frame as given in eq.(6) because of the low luminosity of accretion disks in BL Lacs \cite{Mucke01}.

The number of injected electrons (positrons) per unit volume and time in AGN blob with a Lorentz factor $\gamma_e'$ coming from $\gamma \gamma$ pair production of high-energy photons as given by Aharonian,Atoian \& Nagapetian (1983) \cite{Aharonian83} reads

\begin{eqnarray}
Q_{e,\gamma\gamma}'(\gamma_e') = \frac{3 \sigma_T c}{32} \int_{\gamma_e'}^\infty d\epsilon_{\gamma}' \frac{n'_{\gamma}(\epsilon_{\gamma}')}{\epsilon_{\gamma}'^3} \int_{\frac{\epsilon_{\gamma}'}{4\gamma_e'(\epsilon_{\gamma}'-\gamma_e')}}^\infty d\epsilon_{j}' \frac{n_{j}'(\epsilon_{j}')}{\epsilon_{j}'^2}  \nonumber  \\
\times \left[ \frac{4\epsilon_{\gamma}'^2}{\gamma_e'(\epsilon_{\gamma}'-\gamma_e')}\ln\left( \frac{4\gamma_e'\epsilon_{\gamma}'(\epsilon_{\gamma}'-\gamma_e')}{\epsilon_{\gamma}'}\right) - 8\epsilon_{\gamma}'\epsilon_{j}' \right.  \nonumber   \\
\left. + \frac{2\epsilon_{\gamma}'^2(\epsilon_{\gamma}'\epsilon_{j}'-1)}{\gamma_e'(\epsilon_{\gamma}'-\gamma_e')}-\left( 1-\frac{1}{\epsilon_{\gamma}'\epsilon_{j}'}\right)\left(\frac{\epsilon_{\gamma}'^2}{\gamma_e'(\epsilon_{\gamma}'-\gamma_e')}\right)^2   \right]  \;
\end{eqnarray}
where $n'_{\gamma}(\epsilon'_{\gamma}) = (R_b'/c)Q_{\gamma,pp}'$ is the number density of photons of high energy $\epsilon_{\gamma}'$.

The high-energy injected electrons/positrons ($Q_{e}'$) including both those ($Q_{e,\gamma\gamma}'$), produced in $\gamma\gamma$ pair production and those ($Q_{e,\pi}'$), created directly due to the decay of $\pi^{\pm}$ mesons produced in $pp$ interaction (using eq.(9)) will initiate EM cascades in the AGN blob via the synchrotron radiation, the IC scattering.

In order to determine the stationary state of the population of produced electron distribution $N_e'(\gamma_e')$, the injection function $Q_e'(\gamma_e')$ has been used as a source term in the continuity equation for electrons as given by \cite{Cerruti15}

\begin{eqnarray}
\frac{\partial}{\partial t} \left[N_e'(\gamma_e')\right] = \frac{\partial}{\partial \gamma_e'}\left[\gamma_e'\frac{N_e'(\gamma_e')}{\tau_c(\gamma_e')}\right] + Q_e'(\gamma_e') - \frac{N_e'(\gamma_e')}{\tau_{ad}},
\end{eqnarray}
where we consider the adiabatic time-scale as $\tau_{ad} = 2R_b'/c$. The radiative cooling time, considering both inverse Compton losses and synchrotron losses is given by \cite{Cerruti15}

\begin{eqnarray}
\tau_c(\gamma_e') = \frac{3m_e c}{4(u_B' + u_{ph}')\sigma_T}\frac{1}{\gamma_e'}
\end{eqnarray}
where $u_{ph}'$ is the energy density of photons in co-moving jet frame in equilibrium. 

Using the integral expression given by Inoue \& Takahara (1996) \cite{Inoue96}, the solution of eq. (14) i.e, the cascade electron distribution in stationary state can be evaluated as  

\begin{eqnarray}
N_e'(\gamma_e') = e^{-\gamma_e^*/\gamma_e'}\frac{\gamma_e^*\tau_{ad}}{\gamma_e'^2}\int_{\gamma_e'}^\infty d\zeta Q_e'(\zeta)e^{+\gamma_e^*/\zeta}
\end{eqnarray}
where
\begin{eqnarray}
\gamma_e^* = \frac{3m_e c^2}{8(u_B' + u_{ph}')\sigma_T R_b'}
\end{eqnarray}
indicating the Lorentz factor of electron when $\tau_c(\gamma_e') = \tau_{ad}$. Once the equilibrium pair distribution $N_e'(\gamma_e')$ is known, the associated stationary synchrotron emission is evaluated using Equation (3) and hence found the observable photon spectrum using eq. (11). 

Let $Q_{\gamma,esc}'(\epsilon_{\gamma}')$ be the total gamma ray emissivity from the blob of AGN jet including all processes stated above i.e, the synchrotron and the IC radiation of relativistic electrons, the gamma rays produced in $pp$ interaction and also the synchrotron photons of EM cascade electrons. The observable differential flux of gamma rays reaching at the earth, therefore, can be written as   

\begin{eqnarray}
%\scriptsize
E_{\gamma}^2\frac{d\Phi_{\gamma}}{dE_{\gamma}} = \frac{V'\delta^2\Gamma_j^2}{4\pi d_L^{2}}\frac{E_{\gamma}'^2}{m_e c^2} Q_{\gamma,esc}'(\epsilon_{\gamma}') . e^{-\tau_{\gamma\gamma}^{EBL}}
\end{eqnarray}
where $E_{\gamma} = \delta E_{\gamma}'/(1+z) $ \cite{Atoyan03} relates photon energies in the observer and co-moving jet frame of red shift parameter $z$ respectively with $E_{\gamma}' = m_e c^2 \epsilon_{\gamma}'$, $V' = \frac{4}{3}\pi R_b'^3$ is the volume of the emission region. Here we employ the Franceschini-Rodighiero-Vaccari (FRV) model \cite{Franceschini08,website1} to find the optical depth $\tau_{\gamma\gamma}^{EBL}(\epsilon_{\gamma}',z)$ for gamma-ray photons due to the absorption by the extragalactic background (EBL) light.

The corresponding flux of muon neutrinos reaching at the earth can be written as 
\begin{eqnarray}
%\scriptsize
E_{\nu}^2\frac{d\Phi_{\nu_{\mu}}}{dE_{\nu}} = \xi.\frac{V'\delta^2\Gamma_j^2}{4\pi d_L^{2}} \frac{E_{\nu}'^2}{m_e c^2}Q_{\nu,pp}'(\epsilon_{\nu}') 
\end{eqnarray}
where $E_{\nu} = \delta E_{\nu}'/(1+z) $ \cite{Atoyan03} relates neutrino energies in the observer and co-moving jet frame respectively and the fraction $\xi = 1/3$ is considered due to neutrino oscillation.

\section{Numerical results and discussion}
In the third catalog of AGNs detected by Fermi-LAT listing 1773 objects \cite{Ackermann15}, TXS 0506+056 is one of the most luminous objects with an average flux of $6.5 (\pm 0.2)\times 10^{-9}$ photons cm$^{-2}$ s$^{-1}$ between 1 GeV and 100 GeV. A high-energy neutrino-induced muon track IceCube-170922A, detected on 22 September 2017, was found to be positionally coincident with the flaring $\gamma-$ray blazar, TXS 0506+056 \cite{IceCube18a}. The coincidence detection probability by chance was found to be disfavored at a $3\sigma$ confidence level mainly due to the precise determination of the direction of neutrino \cite{IceCube18a} although no additional excess of neutrinos was found from the direction of TXS 0506+056 near the time of the alert. Assuming a spectral index of −2.13 (−2.0) for the diffuse astrophysical muon neutrino spectrum \cite{Aartsen14}, the most probable energy of the neutrino event was found to be 290 TeV (311 TeV) with the 90\% C.L. lower and upper limits being 183 TeV (200 TeV) and 4.3 PeV (7.5 PeV), respectively \cite{IceCube18a, Ansoldi18}. 
%IceCube collaboration et al. (2018) \cite{IceCube18a} reported that the average integrated muon neutrino energy flux would be $1.8\times 10^{-10}$ erg.cm$^{-2}$.s$^{-1}$ over the energy range 200 TeV to 7.5 PeV for a source that emits neutrinos only during the $\sim 6$-month period corresponding to the duration of the high-energy $\gamma$-ray flare. 
Extensive follow-up observations by the Fermi-Large Area Telescope \cite{Tanaka17} in GeV gamma-rays and by the Major Atmospheric Gamma-ray Imaging Cherenkov (MAGIC) \cite{Mirzoyan17} telescopes in very-high-energy (VHE) gamma-rays above 100 GeV, revealed TXS 0506+056 to be active in all EM bands. The redshift of the blazar has been recently measured to be $z = 0.3365$ \cite{Paiano18} and the luminosity distance, estimated with a consensus cosmology is $d_L \sim 1750$ Mpc \cite{Keivani18}.

The gamma ray variability time scale is found as $t_{ver} \le 10^5$ s by analyzing the X-ray and gamma-ray light curves \cite{Keivani18}. Consequently to describe the electromagnetic SED of TXS 0506+056 over the optical to gamma ray energy range we have chosen the size of emission region of $R_b' = 2.2\times10^{16}$ cm with Doppler boosting factor $\delta = 20$ and bulk Lorentz factor of AGN jet $\Gamma_j = 10.4$  which are strongly consistent with the size inferred from the variability, namely $R_b' \lesssim \delta c t_{ver}/(1+z) \simeq 4.5\times10^{16} (\delta/20) (t_{ver}/10^5 s)$ cm \cite{Keivani18}. 

The low energy part of the experimental EM SED data can be explained well by synchrotron emission of primary relativistic electron's distribution obeying a broken power law as given by Eq.(1) with spectral indices $\alpha_1 = 1.71$ and $\alpha_2 = 4.3$ respectively before and after the spectral break Lorentz factor $\gamma_b' = 8.5\times 10^{3}$. The required kinematic power of relativistic electrons in blazar jet as given by Eq. (2) and the magnetic field to fit the observed data are $L_e' = 2.3\times 10^{42}$ erg/s and $B' = 0.38$ G respectively. Here we have not included the self-absorption of synchrotron photons spectrum. When the self-absorption mechanism \cite{Katarzynski01} is included, the resultant spectrum will show slight mismatch with the observed photon flux at radio energies, particularly VLA and OVRO data.

The inverse Compton scattering of primary electrons with the target synchrotron photons (also including high energy photons) co-moving with the AGN jet as given by Eq. (4) are also found to produce lower part of high energy bump of EM spectrum, particularly from NuSTAR experimental data upto Fermi-LAT data.  The number of `hot' electrons in blob of the AGN jet are estimated to be $n_{e,h}' = 1.7\times10^3$ particles/cm$^{3}$ which is required to produce the EM SED due to both synchrotron and inverse Compton emission. But the acceleration efficiency of electrons in AGN jet may be quite low and it can be assumed to be $\chi_e \approx 10^{-3}$ \cite{Bykov96,Eichler05,Vazza15} due to strong synchrotron and Inverse Compton cooling at relativistic energies and the total number of electrons including `cold' electrons are found out to be $n_e = 1.7\times10^6$ particles/cm$^{3}$. 

In the original proton blazar model high energy gamma rays are produced through synchrotron radiation by high energy protons in strong magnetic field environment. However, due to low magnetic field strength of the source (obtained from the fitting of low energy hump of SED) the gamma ray spectrum of the source can not be modeled with the proton synchrotron radiation. The proton-photon interaction is also found inefficient in the present case  due to low amplitude of target synchrotron photon field. Instead required gamma rays are found to produce in interactions of relativistic protons with the ambient cold protons in the blob. The observed higher energy part of observed EM SED data, particularly those measured with Fermi-LAT and MAGIC observatory, can be reproduced well by the model as estimated following the best fit Eq. (18). The spectral index of the energy spectrum of AGN accelerated cosmic rays is taken as $\alpha_p = - 2.13$ which is consistent with the best fit spectral slope of the observed astrophysical neutrinos of between 194 TeV and 7.8 PeV by IceCube observatory \cite{Halzen17,Aartsen16}. The required accelerated primary proton injection luminosity is found to be $L_p' = 10^{46}$ erg/s. The cold proton number density in jet turns out to be $1.68\times 10^6$ particles/cm$^3$ under charge neutrality condition which provides sufficient targets for hadronuclear interactions with accelerated relativistic protons. The estimated differential gamma-ray spectrum reaching at Earth from this AGN is shown in Fig. 1 along with the different space and ground based observations. It is clear from the figure that the observed spectrum is correctly reproduced by the model. The detection sensitivity of upcoming gamma-ray experiments like the Cherenkov telescope array (CTA) \cite{Ong17} and the Large High Altitude Air Shower Observatory (LHAASO) \cite{Liu17} are also shown in the figure which suggest that these experiments will be able to detect gamma rays up to nearly 100 TeV for any similar kind of event if detected in future and thereby will be able to provide a better understanding of the emission processes.

\begin{figure}[h]
  \begin{center}
% \scalebox{2.5}{
  \includegraphics[width = 0.5\textwidth,height = 0.45\textwidth,angle=0]{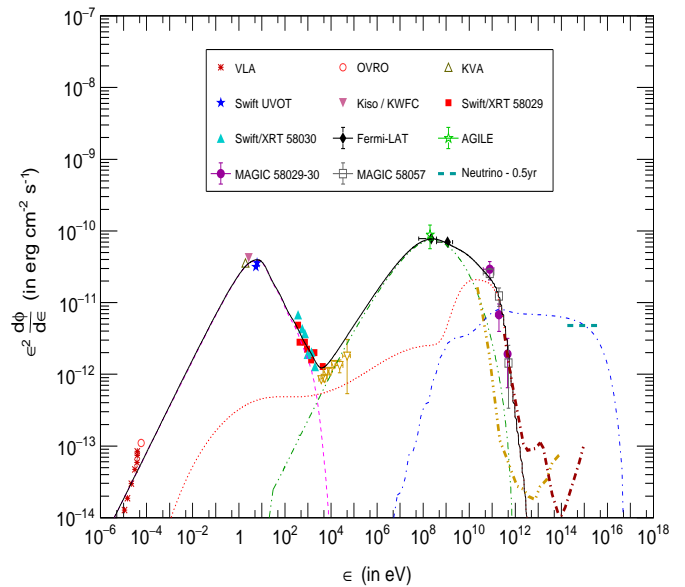}
\end{center}
%  \captionsetup{margin=50pt,font=small,labelfont=bf}
\caption{Estimated differential energy spectrum of gamma rays and neutrinos reaching at the Earth from the blazar TXS 0506+056. The pink small dashed line indicates the low energy component of EM spectrum due to synchrotron emission of relativistic electrons. The green long dash-double-dotted denotes the gamma ray flux produced due to Inverse Compton emission of relativistic electrons in seed photon distribution in co-moving jet frame. The red dotted line represents the gamma ray flux produced from neutral pion decay in $pp$-interaction together with the cascade emission of electron/positron produced in (pionic) $\gamma\gamma$-absorption. The black continuous line represents the estimated overall differential multi wave-length EM SED. The blue small dash-single dotted line indicates the differential muon neutrino flux reaching at earth. The yellow dash-triple-dotted line and brown long dash-single dotted line denote the detection sensitivity of the CTA detector for 1000 h and the LHAASO detector for 1 year respectively. The cyan long dashed line indicates the expected level\cite{Gao18} and energy range of the neutrino flux reaching at earth  to produce one muon neutrino in IceCube in 0.5 year, as observed.}  
\label{Fig:1}
\end{figure}

High energy neutrinos are produced together with gamma rays in pp interactions. The high energy neutrino flux at the Earth from the blazar TXS 0506+056 in active state has also been estimated following Eq. (19) and also shown in Fig. 1 along with the concerned Icecube result. For estimation of the neutrino flux no additional adjustable parameters were available; the same parameters used to describe gamma ray spectrum lead the neutrino flux. The total mechanical jet power of the blazar in jet frame is found out to be $L_{jet}' = 1.2\times 10^{47}$ erg/s and physical jet power after Lorentz boost is $L_{jet} = 1.3\times 10^{49}$ erg/s. It is noticed that $\eta'_p = L_p'/L_{jet} = 8.5\%$ under assumption of electron injection efficiency about $\chi_e \approx 10^{-3}$ i.e, cosmic ray protons carries 8.5\% energy of total jet power in co-moving jet frame which is generally expected acceleration efficiency of cosmic rays in astrophysical sources \cite{Banik17b,Sahakyan18}. 
%   new --------- 
The total jet power in the form of magnetic field and relativistic electron and proton kinetic energy calculated as \cite{Ansoldi18} $L^k_{jet} = \Gamma_j^2 \beta_j c \pi R_b'^2\left[u_e' + u_p' + u_B' \right]$ and found out to be $10^{48}$ erg/s. The estimated kinetic jet power of the blazar is consistent with the Eddington luminosity of $L_{edd}\gtrsim 1.3\times 10^{48}$ erg/s if we assume a super-massive black hole of mass $M_{bh} \gtrsim 10^{10} M_{\odot}$, like blazar S5 $0014+813$ \cite{Ghisellini09}. However, jet power may exceed the Eddington luminosity during outbursts or for a collimated outflow in a jet because in such situations the jet does not interfere with the accretion flow. Note that a moderate excess of jet power over the Eddington luminosity (within a factor of ten) seems physically viable \cite{Gao18,Sadowski15}.
%   new --------- 

The MAGIC collaboration reported the prominent spectral steepening observed gamma ray spectra from the said blazar above $\sim 100$ GeV which confirms the internal $\gamma\gamma$ absorption that is robustly expected as a consequence of $pp$ production of a $\sim 290$ TeV neutrino and also restrict the $\delta$ to a low value. The cascade emission of electron/positron pairs induced by protons has been estimated following Eq. ($14-16$) where we include the contribution of high energy photons along with synchrotron photons in jet frame as target for internal $\gamma\gamma$ absorption of firstly produced gamma rays in $pp$ interaction. This mechanism also found to contribute significantly in the hard X-ray to VHE gamma-ray bands. A primary cosmic ray proton spectrum up to $E_{p,max}' = 10$ PeV in jet frame and magnetic fields of $B' = 0.38$ G which is mutually consistent with the synchrotron radiation of electrons for lower bump in EM SED, can somewhat describe the observed gamma-ray spectra. 

The number of expected muon neutrino event in time $\tau$ can be found from the relation $N_{\nu_{\mu}} = \tau \int_{\epsilon_{\nu,min}}^{\epsilon_{\nu,max}} A_{eff}(\epsilon_{\nu}). \frac{d\phi_{\nu_{\mu}}}{d\epsilon_{\nu}} d\epsilon_{\nu}$ where $A_{eff}$ be the IceCube detector effective area at the declination of the TXS 0506+056 in the sky \cite{IceCube18b,Padovani18,Albert18}. We found that the expected muon neutrino event in IceCube detector from the blazar in 200 TeV and 7.5 PeV energy range are about $N_{\nu_{\mu}} = 1.007$ events in 0.5 years for the flaring VHE emission state with $E_{p,max}' = 10$ PeV. The expected muon neutrino event are about $N_{\nu_{\mu}} = 2.6$ for the same scenario but in the energy range of 32 TeV and 3.6 PeV with $E_{p,max}' = 10$ PeV which is in good agreement with the effective energy range \cite{IceCube18b} of IceCube for astrophysical neutrinos. The model fitting parameters to match the EM SED as well as muon neutrino event are summarized in Table~\ref{table1}.

\begin{table}[h]
  \begin{center}
    \caption{Model fitting parameters for TXS 0506+056 according to proton blazar model.}
    \label{table1}
    \begin{tabular}{c|c}
      \toprule
      Parameters &  Values   \\ \hline
      $\delta$              &  $20$ \\
      $\Gamma_j$              &   $10.4$  \\
      $\theta$              &   $1^{0}$  \\
       $z$                  &  $0.3365$   \\
      $R_b'$  (in cm)            &  $2.2\times 10^{16}$ \\
       $B$ (in G)          &  $0.38$  \\
       $u_B$ (in erg/cm$^{3}$)          &  $5.75\times10^{-3}$  \\
       $\alpha_1$           &   $- 1.71$  \\ 
       $\alpha_2$           &   $- 4.3$  \\
       $\gamma_b'$  &  $8.5\times 10^{3}$ \\
       $\gamma_{e,min}'$  &  $1$ \\
       $\gamma_{e,max}'$  &  $1.5\times 10^{5}$ \\
       $u_e$ (in erg/cm$^{3}$)          &  $4.5\times10^{-2}$  \\
       $L_e'$ (in erg/s)    & $2.3\times 10^{42}$  \\ 
       $n_H$  (in cm$^{-3}$) &  $1.68\times 10^6$  \\
       $\alpha_p$           &   $- 2.13$  \\
       $E_{p,max}'$ (in eV) &  $ 10^{16}$  \\ 
       $L_p'$ (in erg/s)    &  $10^{46}$  \\
       $L^k_{jet}$ (in erg/s)   &  $ 10^{48}$ \\
       $N_{\nu_{\mu}}$   &  1.007  \\ \hline    \hline
    \end{tabular}
  \end{center}
\end{table}

The VHE gamma-ray flux from the blazar is found to be variable i.e, increasing by a factor of up to $\sim 6$ within one day from low state (quiescent state) to the flaring state. The flux variability found mainly in high energy component but not in lower bump of EM spectra from the source disfavors the inverse-Compton origin for such variabilities. There may be two possible scenarios for such variabilities$-$ i) The VHE gamma-ray flux in low state is leptonic in origin, i.e, via inverse-Compton emission from electrons up-scattering synchrotron photons (synchrotron-self-Compton scenario, SSC \cite{Maraschi92,Bloom96,Mastichiadis97}) or photons from the ambient fields (external inverse-Compton, EIC \cite{Dermer92,Dermer93}) but consequently no neutrinos will produce. The higher flux of gamma rays in flaring state can be interpreted when the blazar jet meets with the external cloud \cite{Aharonian17,Barkov12,Dar97} which will provide sufficient target matter (protons) for interaction with accelerated cosmic rays to produce observed high energy gamma rays and neutrinos efficiently. ii) The VHE gamma-ray flux in both low state and flaring state can be explained in a hadronic interaction model using a proton blazar model. In this scenario, observed gamma ray flux can be explained well with hadronic $pp$ interaction of accelerated cosmic rays of comparatively harder spectral slope ($\sim 2.28$) and lowering the maximum energy of accelerated cosmic rays with ambient `cold' proton (in charge neutrality condition with co-accelerated electrons) in low state of the blazar compared to flaring state and subsequently produce neutrinos (of event $N_{\nu_{\mu}} = 0.13$ in 0.5 year) as well.  

\section{Conclusion}
The  coincident detection of the neutrino event, IceCube-170922A with the gamma ray flaring blazar, TXS 0506+056 provide support to the acceleration of cosmic rays in the blazar jet in diffusive shock acceleration process \cite{IceCube18a}. In the framework of proton blazar model, our findings suggest that relative contributions to the total jet power of cold protons, accelerated protons, magnetic field, and accelerated electrons, obtained on the basis of charge neutrality, can explain both the low and high energy bump of the multi-wavelength EM SED and also the observed neutrino event, IceCube-170922A from the flaring blazar, TXS 0506+056 consistently. We find that maximum energy of cosmic ray particle achievable in the blazar is one order less then the ankle energy of cosmic ray energy spectrum or $2\times 10^{17}$ eV in observer frame, is required to explain consistently the observed gamma ray and the neutrino signal from the source. The upcoming gamma-ray experiments like CTA \cite{Ong17} and LHAASO \cite{Liu17}, which are much sensitive up to 100 TeV energies, may provide clearer picture regarding the physical origin of gamma rays if more events like TXS 0506+056 are detected in future.

The gamma ray flux in the quiescent state of the source TXS 0506+056 is smaller by an order or so. Such a fact disfavor the cloud-jet interaction model as in the absence of cloud the gamma ray flux should decrease substantially. One may argue that the quiescent state gamma ray flux is due to inverse Compton process by relativistic electrons. But a fine tuning is needed to produce the exactly same kind of shape and same peak position of the second hump of EM SED in both enhanced and quiescent state if two different processes (hadronic and inverse Compton) are invoked to explain the observations. Recently, IceCube collaboration re-analyzed their historical data and reported significantly an evidence for a flare of 13 muon-neutrino events in the direction of TXS 0506+056 between September 2014 and March 2015 \cite{IceCube18a}. Surprisingly, the blazar TXS 0506+056 was found to be in the quiescent state of both the radio and GeV emission at the arrival time window of such a neutrino flare \cite{Padovani18}. Such an observation favors hadronic interaction mechanism for the production of observed high energy gamma rays and as well as neutrinos for both low and flaring state of the blazar. More elaborate studies are required to understand the production mechanism of the muon-neutrino events from TXS 0506+056 in the quiescent state.

\section*{Acknowledgments}
The authors would like to thank an anonymous reviewer for insightful comments and very useful suggestions that helped us to improve and correct the manuscript.


\begin{thebibliography}{References}
\bibitem{IceCube18a} The IceCube Collaboration et al., Science 361, eaat1378 (2018).
\bibitem{IceCube18b} IceCube Collaboration et al., Science 361, 147 (2018), arXiv:1807.08794.
\bibitem{Ansoldi18} S. Ansoldi et al., Astrophys. J. 863, L10 (2018).
\bibitem{Tanaka17} Y. T. Tanaka, S. Buson, D. Kocevski, The Astronomer's Telegram 10791 (2017).
\bibitem{Keivani18} A. Keivani et al., Astrophys.J. 864, 84 (2018).
\bibitem{Padovani19} P. Padovani et al. Mon. Not. R. Astron. Soc., 484, L104 (2019).
\bibitem{Gao18} S. Gao, A. Fedynitch, W. Winter, and M. Pohl, Nature Astronomy 3,88 (2019), arXiv:1807.04275.
\bibitem{Cerruti19} M. Cerruti, A. Zech, C. Boisson, G. Emery, S. Inoue and J.-P. Lenain, Mon. Not. R. Astron. Soc. 483, L12 (2019).
\bibitem{Liu18} R. Liu et al., ArXiv e-prints (2018), arXiv:1807.05113.
\bibitem{Sahakyan18} N. Sahakyan, Astrophys. J., 866, 109 (2018).
\bibitem{Aharonian00} F.A. Aharonian, New Astronomy 5, 377 (2000).
\bibitem{Morrison84} P. Morrison, D. Roberts \& A. Sadun, ApJ, 280, 483 (1984).
\bibitem{Dar97} A. Dar \& A. Laor, Astrophys. J., 478, L5 (1997).
\bibitem{Hirotani05} K. Hirotani, Astrophys. J. 619, 73 (2005).
\bibitem{Kino04} M. Kino \& F. Takahara, Mon. Not. R. Astron. Soc. 349, 336 (2004).
\bibitem{Celotti93} A. Celotti and A.C. Fabian,  Mon. Not. R. Astron. Soc. 264, 228 (1993).
\bibitem{Rawlings91} S. Rawlings, and R.Saunders, Nature, 349, 138 (1991).
\bibitem{Ghisellini10} G. Ghisellini et al., Mon. Not. R. Astron. Soc. 402, 497 (2010).
\bibitem{Protheroe01} R.J. Protheroe and A. Mucke, ASP Conference Series, vol. 250, (2001).  
\bibitem{Mucke01}  A. M$\ddot{u}$cke and R.J. Protheroe, Astroparticle Physics 15, 121 (2001).
\bibitem{Aharonian17} F. A. Aharonian, M. V. Barkov and D. Khangulyan, Astrophys. J. 841, 61, (2017).
\bibitem{Petropoulou15} M. Petropoulou and A. Mastichiadis, Mon. Not. R. Astron. Soc. 447, 36 (2015).
\bibitem{Katarzynski01} K. Katarzy\'{n}ski, H.Sol, and A. Kus, A\&A 367, 809 (2001).
\bibitem{Bottcher13} M. B$\ddot{o}$ttcher, A. Reimer, K. Sweeney, and A. Prakash, Astrophys. J. 768, 54 (2013).
\bibitem{Bykov96} A. M. Bykov \& P. M\'{e}sz\'{a}ros, Astrophys. J. 461, L37 (1996).
\bibitem{Eichler05} David Eichler \& Eli Waxman, Astrophys. J., 627, 861, (2005).
\bibitem{Vazza15} F. Vazza, D. Eckert, M. Br$\ddot{u}$ggen and B. Huber, Mon. Not. R. Astron. Soc. 451, 2198 (2015).
\bibitem{Blumenthal70} G. R. Blumenthal \& R. J. Gould, Rev. Mod. Phys., 42, 237 (1970).
\bibitem{Inoue96} S. Inoue \& F. Takahara, Astrophys. J., 463, 555 (1996).
\bibitem{Jones68} F. C. Jones, Phys. Rev., 167, 1159 (1968).
\bibitem{Dermer02} C. D. Dermer and R. Schlickeiser, Astrophys. J., 575, 667 (2002).
\bibitem{Atoyan03} A. M. Atoyan, C. D. Dermer, Astrophys. J. 586, 79 (2003).
\bibitem{Malkov01} M. A. Malkov and L. O. Drury, Rep. Prog. Phys. 64, 429 (2001).
\bibitem{Cerruti15} M. Cerruti, A. Zech, C. Boisson and S. Inoue, Mon. Not. R. Astron. Soc. 448, 910 (2015).
\bibitem{Anchordoqui07} L. A. Anchordoqui, J. F. Beacom, H. Goldberg, S. Palomares-Ruiz, and T. J. Weiler, Phys. Rev. D 75, 063001 (2007).
\bibitem{Banik17a} P. Banik, B. Bijay, S. K. Sarkar, and A. Bhadra, Phys. Rev. D 95, 063014 (2017a).
\bibitem{Kelner06} S. R. Kelner, F. A. Aharonian, and V. V. Bugayov, Phys. Rev. D 74, 034018 (2006).
\bibitem{Banik17b} Prabir Banik and Arunava Bhadra, Phys. Rev. D 95, 123014 (2017b)
\bibitem{Aharonian08} F. A. Aharonian, D. Khangulyan \& L. Costamante, Mon. Not. R. Astron. Soc. 387, 1206 (2008).
\bibitem{Aharonian83} F. A. Aharonian, A. M. Atoian\& A. M. Nagapetian, Astrofizika 19, 323 (1983).
\bibitem{Inoue96} S. Inoue, F. Takahara, Astrophys. J. 463, 555 (1996).
\bibitem{Franceschini08} A. Franceschini, G. Rodighiero, and M. Vaccari, Astron. Astrophys. 487, 837 (2008).
\bibitem{website1} http://www.astro.unipd.it/background/. 
\bibitem{Ackermann15} M. Ackermann et al., Astrophys. J. 810, 14 (2015).
\bibitem{Aartsen14} M. G. Aartsen, M. Ackermann, J. Adams et al., Astrophys. J. 796, 109 (2014).
\bibitem{Mirzoyan17} R. Mirzoyan, Astronomerʼs Telegram, 10817, 1 (2017).
\bibitem{Paiano18} S. Paiano, R. Falomo, A. Treves \& R. Scarpa, ApJL 854, L32 (2018). 
\bibitem{Halzen17} F. Halzen, Nature Physics, 13, 232 (2017).
\bibitem{Aartsen16} M. G. Aartsen, Astrophysical J., 833, 3 (2016).
\bibitem{Ong17} R.A. Ong, PoS, ICRC2017 1071 (2017).
\bibitem{Liu17} C. Liu, for the LHAASO Collaboration, PoS, ICRC2017 424 (2017).
%
\bibitem{Ghisellini09} G. Ghisellini, L. Foschini, M. Volonteri, G. Ghirlanda, F. Haardt, D. Burlon4 and F. Tavecchio, Mon. Not. R. Astron. Soc.399,L24 (2009)
\bibitem{Sadowski15} A. Sadowski \& R. Narayan, Mon. Not. R. Astron. Soc. 453, 3213 (2015).
%
\bibitem{Padovani18} P. Padovani, P. Giommi, E. Resconi, T. Glauch, B. Arsioli, N. Sahakyan, and M. Huber, Mon. Not. R. Astron. Soc. 480, 192 (2018).
\bibitem{Albert18} A. Albert et al., Mon. Not. R. Astron. Soc. 482, 184 (2019).
\bibitem{Maraschi92} L. Maraschi, G. Ghisellini, A. Celotti, Astrophys. J., 397, L5 (1992).
\bibitem{Bloom96} S. D. Bloom, A. P. Marscher, Astrophys. J., 461, 657 (1996).
\bibitem{Mastichiadis97} A. Mastichiadis, J. G. Kirk, A\&A, 320, 19 (1997).
\bibitem{Dermer92} C. D. Dermer, R. Schlickeiser, A. Mastichiadis, A\&A, 256, L27 (1992).
\bibitem{Dermer93} C. D. Dermer, R. Schlickeiser, Astrophys. J., 416, 458 (1993).
\bibitem{Barkov12} M. V. Barkov, V. Bosch-Ramon and F. A. Aharonian, Astrophys. J. 755, 170 (2012).



%\bibitem{Hillas} A. M. Hillas, ARA\&A, 22, 425 (1984).
%\bibitem{Biermann87} P. L.Biermann \& P. A. Strittmatter, ApJ, 322, 643 (1987).
%\bibitem{Fossati98} G. Fossati, L. Maraschi, A. Celotti, A. Comastri, G. Ghisellini, MNRAS, 299, 433 (1998).
%\bibitem{Ulrich97} M.-H. Ulrich, L. Maraschi, C. M. Urry, ARA\&A, 35, 445 (1997).
%\bibitem{Costamante01} L. Costamante et al., A\&A, 371, 512 (2001).
%bibitem{Abdo11} A. A. Abdo et al., ApJ, 736, 131 (2011).

%\bibitem{Mastichiadis05} A. Mastichiadis, R. J. Protheroe, and J. G. Kirk, A\&A 433, 765-776 (2005).
%\bibitem{Sironi13} L. Sironi, A. Spitkovsky, J. Arons, ApJ, 771, 54 (2013).
%\bibitem{Mannheim95} K. Mannheim, Astropart. Phys., 3, 295 1995.
%\bibitem{Halzen97} F. Halzen, E. Zas, ApJ, 488, 669 (1997).
%\bibitem{Mucke03} A. M$\ddot{u}$cke et al., Astropart. Phys., 18, 593 (2003).
%\bibitem{Aartsen13} M. G. Aartsen, R. Abbasi, Y. Abdou et al., Sci, 342, 1242856 (2013).
%\bibitem{Liu18} Ruo-Yu Liu et al., arXiv:1807.05113 (2018). 

%\bibitem{Ghisellini} G. Ghisellini el al., Mon. Not. R. Astron. Soc. 399, L24-L28 (2009). for agn S5 0014 +813




\end{thebibliography}
\end{document}